\documentclass[twocolumn]{aa}
\usepackage{graphicx}
\usepackage{natbib}

\usepackage{txfonts}
\begin{document}
   \title{Discovery of a bipolar X-ray jet from the T Tauri star  DG Tau}


   \author{Manuel G\"udel
          \inst{1,2,3}
          \and
	  Stephen L. Skinner
	  \inst{4}
	  \and
	  Marc Audard
	  \inst{5,6}
	  \and
 	  Kevin R. Briggs
	  \inst{1}
	  \and
 	  Sylvie Cabrit
	  \inst{7}
         }

   \offprints{M. G\"udel}

   \institute{Paul Scherrer Institut, W\"urenlingen and Villigen,
              CH-5232 Villigen PSI, Switzerland \\
              \email{guedel, briggs@astro.phys.ethz.ch}
         \and
             Max-Planck-Institute for Astronomy, K\"onigstuhl 17,  
             69117 Heidelberg, Germany
         \and
             Leiden Observatory, Leiden University, PO Box 9513, 2300 RA Leiden, 
	     The Netherlands 
          \and
	     CASA, 389 UCB,
             University of Colorado,
             Boulder, CO 80309-0389,
             USA; \email{skinners@casa.colorado.edu}  
          \and
	     Integral Science Data Centre, Ch. d'Ecogia 16,
	     CH-1290 Versoix, Switzerland; 
	     \email{marc.audard@obs.unige.ch}
          \and
	     Observatoire de Gen\`eve, University of Geneva, Ch. de Maillettes 51,
	     1290 Sauverny, Switzerland 
	  \and   
	     L'Observatoire de Paris,
             61, avenue de l'Observatoire,
             75014 Paris;
	     \email{sylvie.cabrit@obspm.fr}
          }

   \date{Received 2007; accepted 2007}

   \abstract
   {} 
   {We have obtained and analyzed {{\it Chandra}} ACIS-S observations of the strongly accreting classical 
   T Tauri star DG Tau. Our principal goals are to map the immediate environment of the star
   to characterize possible extended X-rays formed in the jet, and to re-visit the anomalous,
   doubly absorbed X-ray spectrum of DG Tau itself. 
    }
   {We combine our new ACIS-S data with a data set previously obtained. The data are superimposed
   to obtain flux and hardness images. Separate X-ray spectra are extracted for DG Tau and areas outside its
   point spread function.
   }
   {We detect a prominent X-ray jet at a position angle of PA $\approx 225$~deg (tentatively suggested by
   \citealt{guedel05}), coincident with the optical jet axis. We also identify a counter jet at
   PA = 45 deg. The X-ray jets are detected out to a distance of $\approx 5\arcsec$ from the star, their sources
   being extended at the ACIS-S resolution. The jet spectra are soft, with a best-fit electron temperature
   of 3.4~MK. We find evidence for excess absorption of the counter jet. The spectrum of the DG Tau
   point source shows two components with largely different temperatures and absorption column densities.
   }
   {The similar temperatures and small absorbing gas columns of the jet sources and the soft component 
   of the ``stellar'' source suggest that these sources are related, produced either by shocks  
   or by magnetic heating in the jets. Cooling estimates suggest that the pressure in the hot gas   
   contributes to jet expansion. The hard ``stellar'' component, on the other hand, is associated with
   a stellar corona or magnetosphere. The excessive photoelectric absorption of this component suggests
   the presence of dust-depleted accretion streams above coronal magnetic fields. }
   \keywords{Stars: coronae --
             Stars: formation --
	     Stars: individual: DG Tau --
             Stars: pre-main sequence --
             Stars: winds, outflows --
             X-rays: stars }

   \maketitle
%

\section{Introduction}

Pre-main sequence stars  
show various signs of accretion and outflow, such as stellar winds inducing 
high mass-loss rates (e.g., \citealt{dupree05, johns07, kwan07}), molecular outflows observed 
in molecular lines (e.g., \citealt{bachiller96}),
and accompanying optical (e.g., \citealt{hirth97, eisloeffel98}) 
and radio jets (e.g., \citealt{anglada95}).  The most evident manifestation of 
outflows are the optically visible jets and their associated Herbig-Haro 
(HH) objects at distances up to several arcminutes from the star. These 
structures are excited by internal shocks or in regions where the fast mass 
stream encounters the interstellar medium and shock-ionizes the gas (for a 
review of Herbig-Haro  flows, see \citealt{reipurth01}). Under ideal 
circumstances (low extinction, strong ionization), optical jets can be identified 
at distances as close as 0$\farcs$1 to the star \citep{bacciotti02}. 
The same compact jets are also routinely detected
at radio wavelengths, where the emission mechanism  is thought 
to be bremsstrahlung from the shock-heated gas \citep{rodriguez95, anglada95}.
Radio brightness temperatures suggest overall gas temperatures of order  $10^4$~K. 
This picture is ambiguous, however, as a number of non-thermal jets have been
suggested from radio polarization or synchrotron-like spectral shapes 
(e.g., \citealt{curiel93, ray97}).
Magnetic fields may thus play a role not only in launching the jets, but in 
their propagation as well.

Outflow processes are prone to producing X-rays,
given that shocks with shock jump velocities of order several  hundred 
km~s$^{-1}$ are possible. 
The relevant theory and a simple model have been discussed  by \citet{raga02}.
The strong-shock temperature can be expressed as 
$T\approx 1.5\times 10^5v_{\rm 100}^2$~K (for fully ionized gas) where 
$v_{100}$ is the shock  speed relative to a target in units of 
100~km~s$^{-1}$. Jet speeds are typically of order $v = 300$--$500$~km~s$^{-1}$ 
\citep{eisloeffel98, anglada95, bally03}, in principle
allowing for shock speeds of similar magnitude. If a flow 
shocks a standing medium at 400~km~s$^{-1}$, then $T \approx 2.4$~MK. 

\begin{table*}[t!]
\caption{Observations}
\label{obslog}
\begin{tabular}{llll}
\hline
            \hline
            \noalign{\smallskip}
            \multicolumn{4}{l}{\it Previous observations:}\\
            \noalign{\smallskip}
            \hline
	    Instrument                 &  {\it Chandra}  ACIS-S          &   {\it XMM-Newton} EPIC         &                        \\ 
	    ObsID                      &  4487			         &  0203540201		           &                        \\ 
	    Start time  (UT)           &  2004-01-11\ 02:58:51	         &  2004-08-17\ 06:08:10           &                        \\ 
	    End time (UT)              &  2004-01-11\ 11:52:21           &  2004-08-17\ 17:32:46           &                        \\ 
	    Exposure time  (s)         &  29717			         &  41076 		           &                        \\ 
            \hline
            \noalign{\smallskip}
            \multicolumn{4}{l}{\it New observations:}\\
            \noalign{\smallskip}
            \hline
	    Instrument                 &  {\it Chandra}  ACIS-S          &   {\it Chandra}  ACIS-S         &  {\it Chandra}  ACIS-S     \\			      
	    ObsID                      &  6409			         &   7247                          &  7246		         \\	  		      
	    Start time  (UT)           &  2005-12-15\ 11:03:14	         &   2005-12-17\ 09:37:05          & 2006-04-12\ 17:13:23 	   \\	  		      
	    End time (UT)              &  2005-12-15\ 16:17:01	         &   2005-12-17\ 14:34:37          & 2006-04-13\ 01:37:19	  \\	  		      
	    Exposure time  (s)         &  16252			         &   15946                         & 27811		         \\	  		      
          \hline
\end{tabular}
\end{table*}

Faint, soft X-ray emission  has been detected from a few protostellar 
HH objects \citep{pravdo01, pravdo04, pravdo05, favata02, bally03, tsujimoto04, grosso06}.
\citet{bally03} used a {\it Chandra} observation to show that  X-rays 
form within an arcsecond of the protostar L1551 IRS-5 while the star itself
is too  heavily obscured to be detected. As this example illustrates, 
the jet-launching region of powerful protostellar jets is 
often inaccessible to optical, near-infrared or X-ray studies due to 
excessive absorption. However, a class of strongly accreting, optically 
revealed classical T Tauri stars (CTTS) also exhibits so-called micro-jets 
visible in optical lines \citep{hirth97}, with flow speeds similar to 
protostellar jets. CTTS micro-jets have  the 
unique advantage that they can - in principle - be followed down to the 
acceleration region close to the star both in the optical and in X-rays. 
For example, \citet{bacciotti00, bacciotti02}  used the HST to trace the 
jet of the CTTS DG Tau  to within 0$\farcs 1$ of the star. 

DG Tau  is a most outstanding T Tauri star for X-ray studies. A {\it Chandra} 
high-resolution X-ray image has shown tentative evidence for the presence of faint X-rays
along the optical jet. Both {\it XMM-Newton} \citep{guedel07b} and {\it Chandra} \citep{guedel05}
low-resolution CCD spectra of DG Tau are anomalous, showing a ``two-absorber X-ray'' (TAX) 
spectrum in which two independent X-ray 
components are each subject to different absorption column densities. 

To study DG Tau's X-ray emission further, we obtained new {\it Chandra} observations 
that, together with the previous observations, result in three times more 
{\it Chandra} exposure than analyzed before.
The present paper describes the new observations, puts them into a context of the 
previous results, and discusses some tentative models for the jet X-ray emission.

\section{The target: DG Tau and its jets}

\subsection{An accreting, jet-driving T Tauri star}

DG Tau is a single classical T Tau star \citep{leinert91}, surrounded by a disk of 
dust \citep{dutrey96, testi02} and gas (\citealt{dutrey96, kitamura96a, kitamura96b} = 
K96a,b; \citealt{testi02}). The dust disk as detected 
in the millimeter continuum is relatively compact (due to limited sensitivity of present-day 
instrumentation to optically thin emission of the outer disk), with a size 
of $1\farcs 1\times 0\farcs 6$  (\citealt{dutrey96}, also K96b). Dust 
is also present at larger distances as inferred from extinction of the counter jet 
(see below). CO gas remains easier to detect at large distances from the star; K96a,b 
find a $^{13}$CO disk size of about $40^{\prime\prime}\times 30^{\prime\prime}$ although we
note that this structure is not a Keplerian disk but more akin to a residual, flattened envelope. 
Interestingly, line shifts as a function of position indicate that the outer parts of 
the disk are {\it expanding}, perhaps driven by interactions between the stellar wind 
and the disk surface. This latter wind would also be responsible for blowing off residual 
envelope gas; a spherical gas envelope is indeed not present anymore, as inferred 
from modeling of the $^{13}$CO line (K96a). 

DG Tau is a strongly accreting CTTS, with an observed accretion rate of 
$\dot{M} = (10^{-7.34}-10^{-6.13})M_{\odot}$~yr$^{-1}$ \citep{white01, white04}. 
It also ejects a well studied collimated jet with knots and bow shocks out to at least 
11$^{\prime\prime}$, with velocities of several 100~km~s$^{-1}$ (e.g., \citealt{lavalley97, 
eisloeffel98, dougados00}). Optical evidence for a counter jet has been reported 
\citep{lavalley97, pyo03}.

\subsection{Previous X-ray observations}

We have previously obtained two short X-ray observations of DG Tau (Table~\ref{obslog}). 
Our previous {\it Chandra} ACIS-S observation \citep{guedel05} showed tentative indications of 
very faint, soft emission along the optical forward jet out to a distance of about 5\arcsec\ to the
SW. A total of 17 excess counts were collected (including the area around the faint counter jet) in
the energy range of 0.4--2.4~keV.

The X-ray spectrum of DG Tau itself revealed  two independent X-ray components in both the 
{\it Chandra} and {\it XMM-Newton} observations \citep{guedel05, guedel07b}; 
a soft, little absorbed component is emitted by cool ($\approx$3--4~MK) plasma; 
and a hard, strongly absorbed component originates from hot, occasionally flaring ($\approx 20$--70~MK) plasma.
The soft component was attributed to emission from the base of the jets.

\section{New Chandra observations}\label{obs}

We have obtained new observations of DG Tau with {\it Chandra} ACIS-S (Table~\ref{obslog}), for
a total of $\approx$60~ks of exposure time. We have merged these data with the previously 
obtained {\it Chandra} data to produce images with an equivalent exposure time of 
approximately 90~ks, i.e., three times as much as reported in \citet{guedel05}. 
The new observations were collected in three segments,
two shorter ones in December 2005 (ObsID 6409 and 7247) and a longer one in April 2006 (ObsID 7246).
The data reduction followed standard {\it Chandra} CIAO analysis 
threads\footnote{http://cxc.harvard.edu/ciao/guides/acis\_data.html; CIAO version 3.3.0.1 was used} 
as described in \citet{guedel05}. We used the ``Very Faint'' mode to efficiently reduce background 
radiation. DG Tau's optical sky coordinates for Epoch 2005.0 are RA(2000.0) = 04h 27m 04.697s,
dec(2000.0) = $+26\deg\ 06\arcmin\ 16\farcs 10$ \citep{ducourant05}.

\begin{figure*}[t!]
\hbox{
\includegraphics[angle=-0,width=9.44cm]{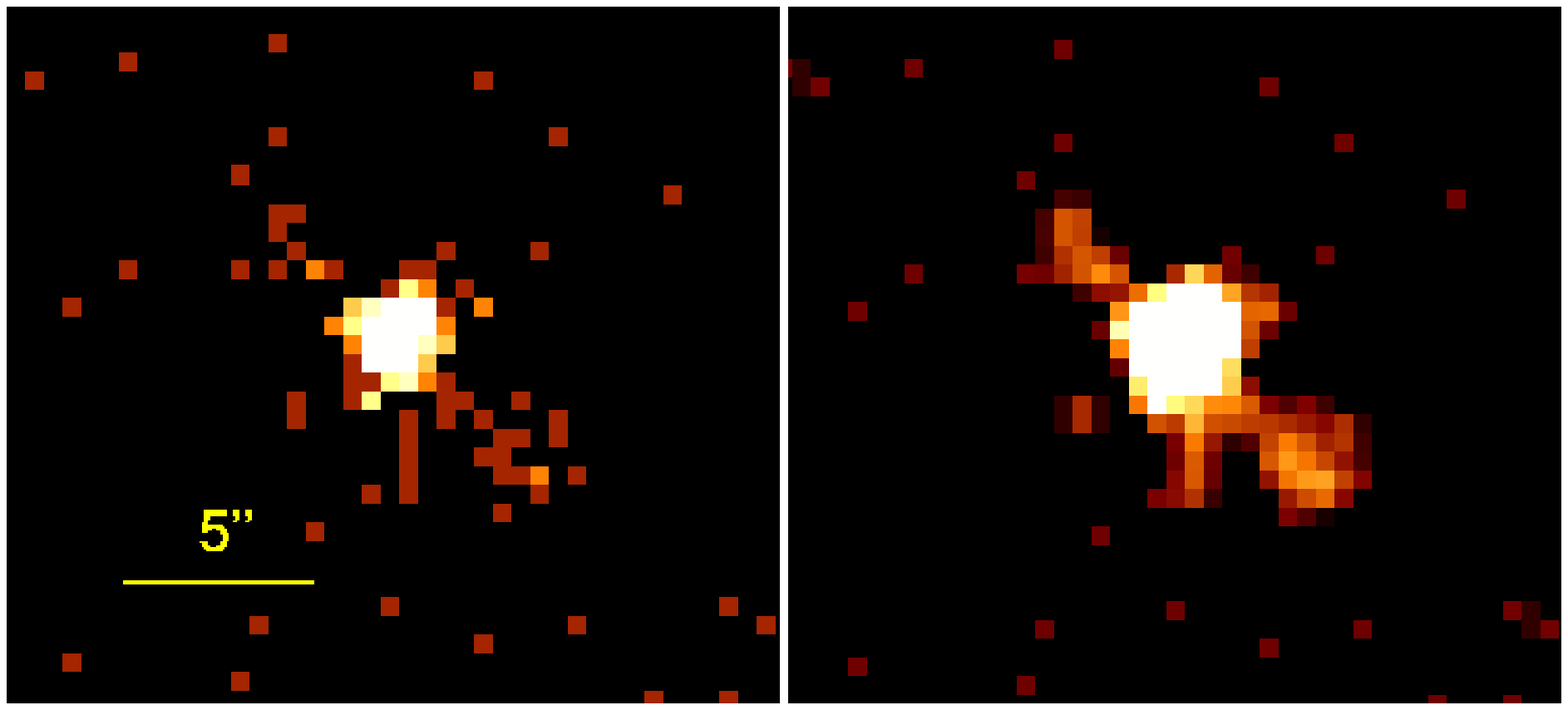}
\hskip -0.04truecm\includegraphics[angle=-0,width=4.44cm]{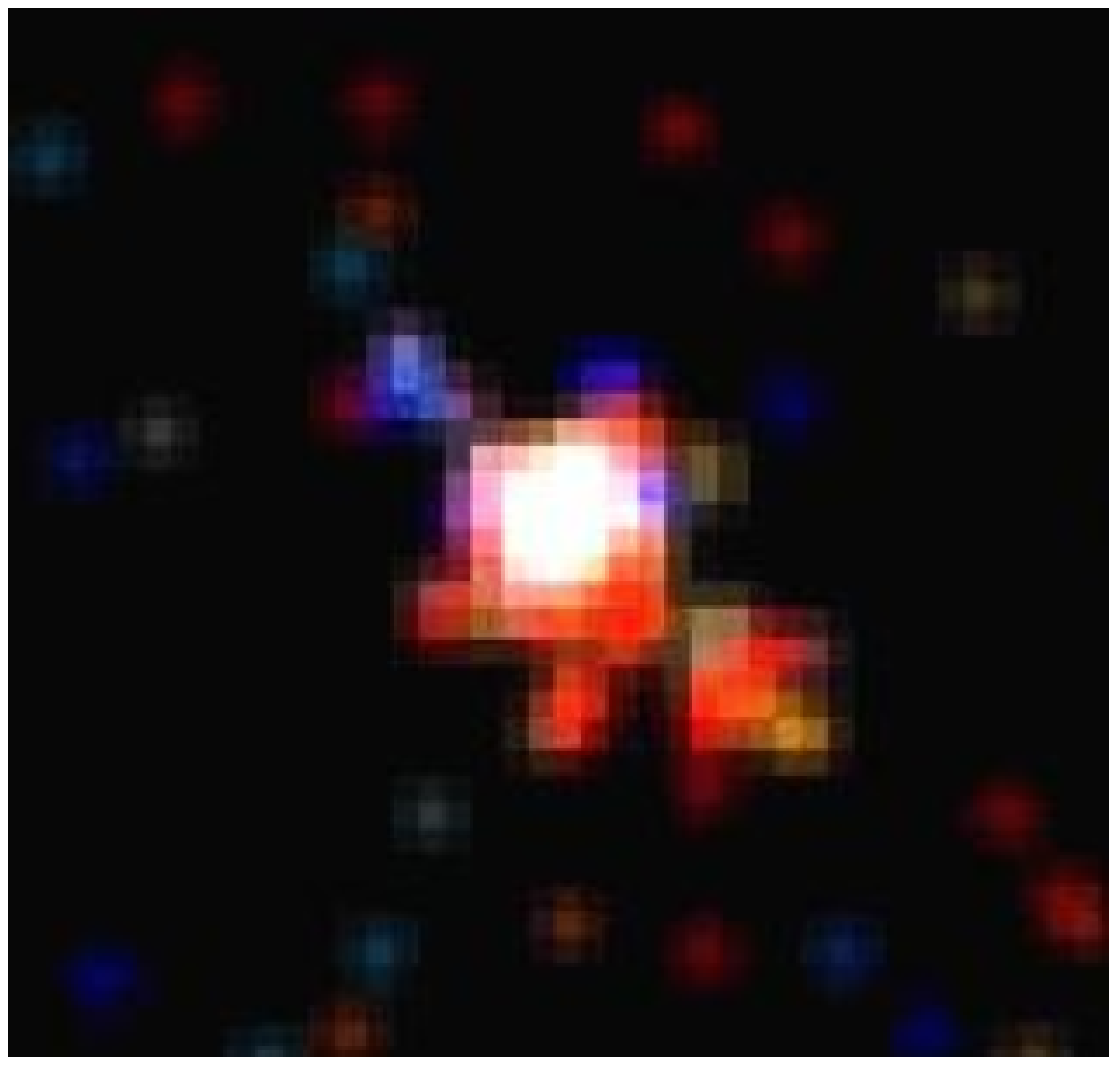}
\hskip -0.05truecm\includegraphics[angle=-0,width=4.48cm]{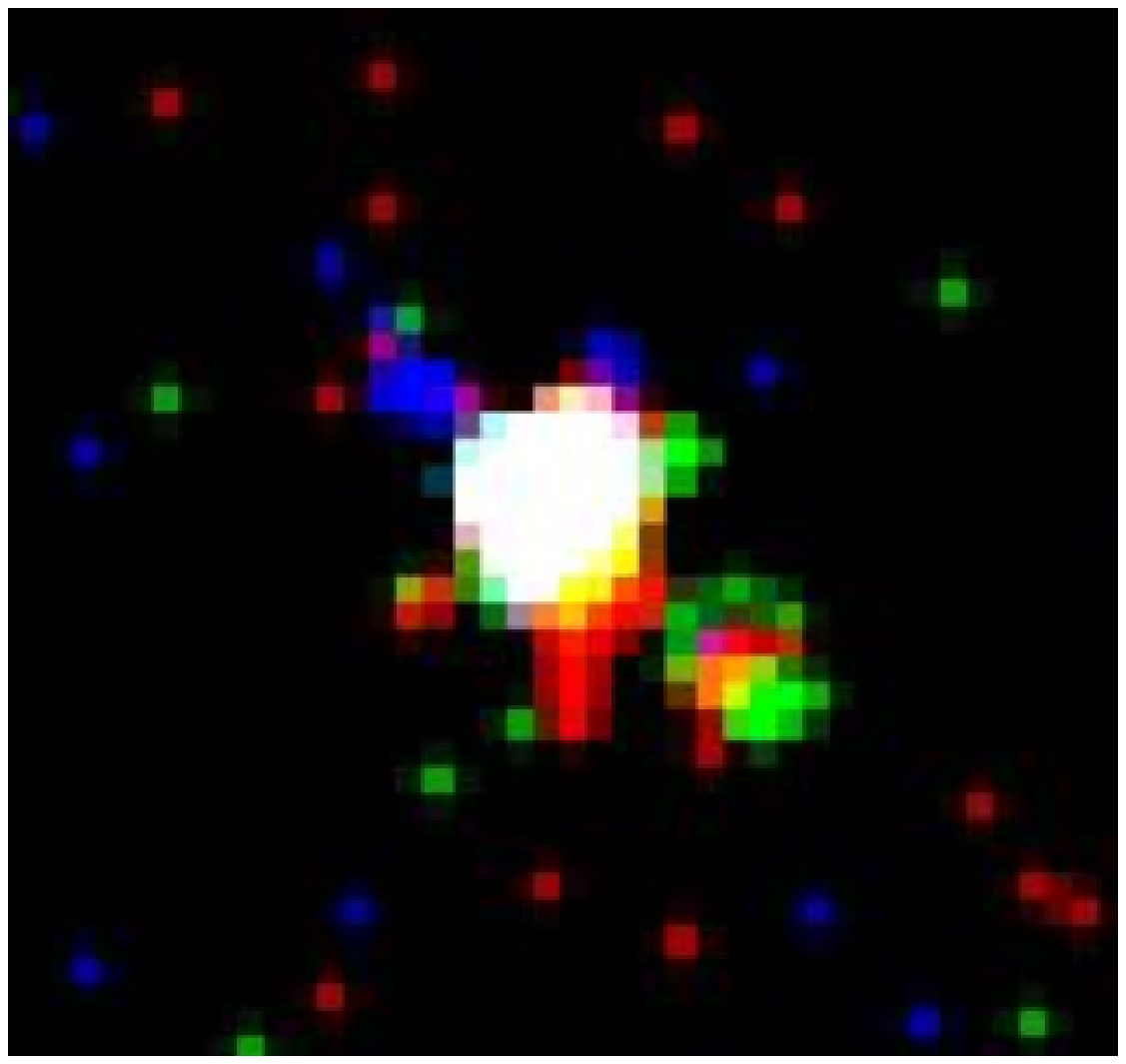}
}
\caption{{\it Chandra} ACIS-S images of DG Tau and its jets. North is up and east is to the left. {\it From left to
right:} {\it a)} Count image for the 0.6-1.7~keV range; {\it b)} same but smoothed; {\it c)} color-coded,
smoothed hardness image; {\it d)} same but with different color coding (see text for details). Pixel size is
0$\farcs$492.}
\label{jetim} 
\end{figure*}
	
We extracted stellar X-ray spectra individually for each observation using counts within 
$1\farcs 4-1\farcs 7$ (depending on the shape of 
the stellar image) of the centroid position of the DG Tau stellar source. This extraction
radius contains approximately 95\% of the power in the PSF (see {\it Chandra} Proposers' Observatory Guide 
[POG] v.9).\footnote{http://cxc.harvard.edu/proposer/POG/html/} Counts from regions 
attributed to the jets were extracted from circles with radii of 1.75\arcsec\ and 2.6\arcsec, 
offset by a few arcseconds to the NE and SW, respectively (see \citealt{guedel05} for an 
illustration).

Spectra were produced for DG Tau, the SW (forward) jet, and the NE (counter) jet for each 
observation (using the CIAO specextract task). We fitted the spectra with simple thermal, 
collisional-ionization equilibrium models (vapec) in the XSPEC software package (version 11.3.1; 
\citealt{arnaud96}), complemented with photoelectric absorption models. DG Tau
required two thermal models that were each subject to different amounts of absorption. The two jet
sources were fitted jointly  (see below). We fixed all element abundances at values that are typical for T Tauri stars,
as used in the XEST project (\citealt{guedel07a, telleschi07a}, and references therein).\footnote{The
adopted abundances are, with respect to the solar photospheric abundances given by \citet{anders89}:
C = 0.45, N = 0.788, O = 0.426, Ne = 0.832, Mg = 0.263, Al = 0.5, Si = 0.309, S = 0.417, Ar = 0.55, Ca = 0.195,
Fe = 0.195, Ni =0.195.}

Because the ACIS-S background level is negligible for our small sources (see below), 
we took advantage of the C statistic \citep{cash79} for the spectral 
fits in XSPEC, applied to unbinned spectra in the energy range of 0.2--7~keV for DG Tau and 0.2-2.2~keV 
for the jet sources.

\section{The X-ray jets of DG Tau}

Our X-ray images (Fig.~\ref{jetim}) provide clear evidence for both a forward (to the SW) and a counter-jet 
(to the NE), symmetrically arranged outside the stellar point-spread function (PSF) out to distances 
of about 5\arcsec. This is the first and so far only young-stellar double X-ray jet that can be followed 
essentially down to the star. We now discuss the spatial and spectral results in detail.

\subsection{Jet morphology}\label{morph}

An X-ray image of the DG Tau environment is shown in Fig.~\ref{jetim}a, with a pixel size
of 0$\farcs$492. This image
was produced by combining counts from all four {\it Chandra} observations. The jet proper motion 
amounts to about $0\farcs 15$--$0\farcs 3$ per year (\citealt{eisloeffel98} for the 1983--1990 time interval;
\citealt{dougados00} for 1994--1997). If the proper motion was similar during the time interval of 
our {\it Chandra} observations (2004--2006),  then the jet features would shift by about one ACIS-S 
pixel between the earliest and the latest observation; such a shift is not critical given the size 
of the jet source (see below). To minimize the extension of the combined image of the DG Tau stellar 
PSF due to slight systematic offsets between  the attitude solutions of the observations,
we determined the stellar centroid coordinates in each exposure using the CIAO wavdetect 
task, and then shifted the centroids to a common coordinate (the maximum shift applied to
one of the exposures was $0\farcs 42$, or less than one pixel).  Also, the
standard ``pixel randomization'' procedure was turned off (but was left on
for the data sets used for spectral extraction). Because
the boresight coordinates are slightly different for the different exposures,
the data were reprojected to a common tangent point, using the reproject\_events
task in CIAO.

To suppress background and to emphasize the soft jet sources, only counts  falling within 
the 0.6--1.7~keV range are plotted in the figure.  There is clear
evidence for a jet-like extension to the SW along a position angle of $\approx$225~deg, but 
we also find  a significant excess of counts in the NE direction (PA $\approx$45~deg). 
This is coincident with the jet optical axis, which for the SW jet has been given as 222~deg 
\citep{lavalley97}, 226~deg \citep{solf93}, or, depending on individual knots, 217-237~deg 
\citep{eisloeffel98}.
We are not aware of any background sources in this region that could produce additional 
X-ray emission. 

Only very few
counts from the jet regions fall outside the $\approx$0.6--1.7~keV energy range; we therefore use this 
range for our statistics (somewhat different from the energy range 0.4--2.4~keV used by \citealt{guedel05}). 
The SW jet contains, within the extraction circle defined here,
7 and 11 counts for the 2004 and the combined 2005/06 observations, respectively (including
one count in the SW jet at 0.58~keV in ObsID 6409). For the NE jet, 
the numbers are 4 and 5 counts, respectively. Based on the 2004 observation, we would 
have expected to detect, in the 2005/06 observations, a total of 14 counts in the SW jet and 8 
counts in the NE jet, but the differences to the actually observed counts are within about
1$\sigma$ of the uncertainties due to counting statistics, and are therefore not significant. 
The combined {\it Chandra} exposures thus  collected 
a total of 18 and 9 counts in the SW and the NE jet, respectively (the nearest counts outside
this energy range were: one count at 0.43~keV in the SW jet of ObsID 4487; and one count at 2.1~keV 
in the SW jet of ObsID 6409). 

We then estimated potential contamination from the diffuse (sky and detector) background and from
the PSF of the bright stellar DG Tau X-ray source. The diffuse background was estimated by
extracting counts from a large area in the vicinity of the DG Tau system and scaling them to
the jet extraction areas. We found the background contribution to be, statistically, 0.6 and 0.3~cts
for the SW and the NE jet area, respectively. We modeled the DG Tau PSF using the MARX 
software.\footnote{http://space.mit.edu/CXC/MARX: version 4.2.0 was used} Boresight coordinates
and roll angles were identical to those of the observations, and the simulated source was put
at the sky coordinates of DG Tau (the simulations 
were done separately for the 2004 and the 2005/06 observations, but the results agreed with each other).
The simulation used the best-fit spectrum of DG Tau except that the flux (or the exposure time)
was much higher. We then extracted 
counts in the PSF wings using the identical extraction regions used for the jets, in the 0.6--1.7~keV
range, and found a statistical contribution of 0.9 and 0.7~cts of the PSF wings to the SW and
NE jet source, respectively. We conclude that one count per jet area is likely to be due to contamination.
This is considerably less than the Poisson uncertainty in our count numbers.
These X-ray jet sources are highly significant (see discussion 
in \citealt{guedel05} for the first exposure only). A linear feature pointing from the stellar PSF 
to the south, however, appears to be due to a coincidental arrangement of only four counts.

\begin{figure}[t!]
\hbox{
\includegraphics[angle=0,width=4.41cm]{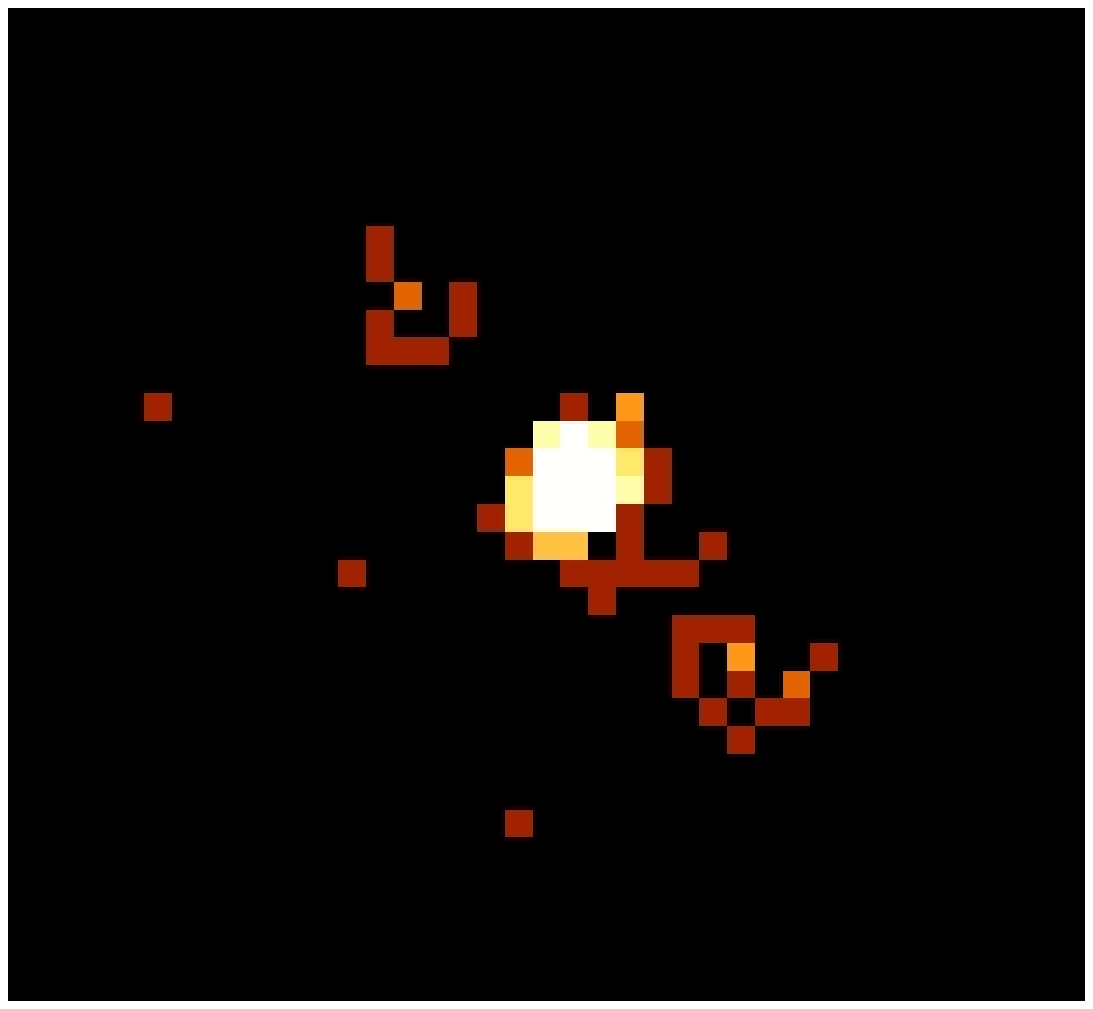}
\includegraphics[angle=0,width=4.4cm]{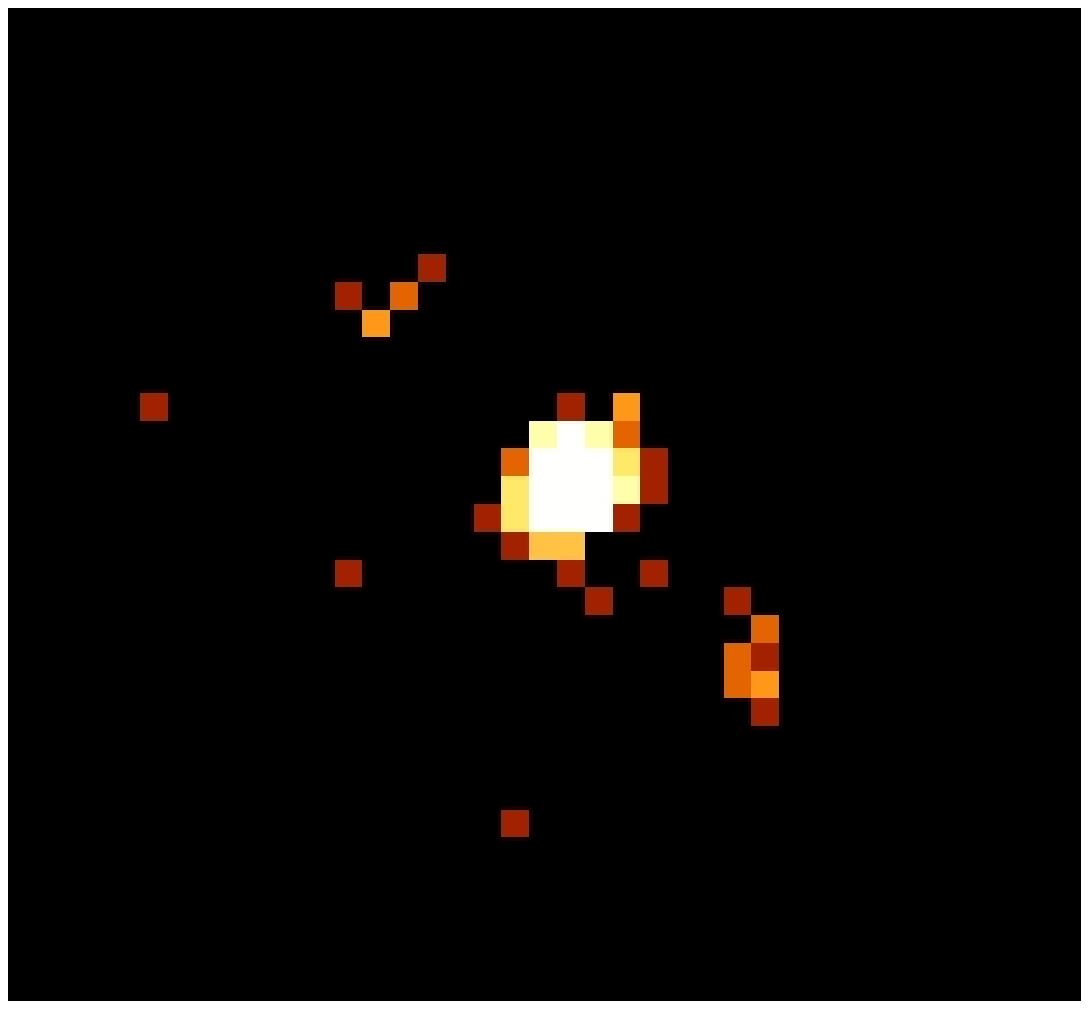}
}
\caption{Comparison of two simulations of the DG Tau X-ray sources. {\it Left:} 
The two jet sources are extended; the NE jet source is represented by a Gaussian source with sigma 
= $0\farcs 8$; the SW jet source  consists of  a Gaussian with sigma = 
$1\arcsec$ plus a linear source at a position angle of 225~deg between star and Gaussian source.
The number of counts in each source is similar to the true observations for an exposure time
of 90~ks. Only counts in the range of 0.6--1.7~keV are shown. -- {\it Right:} Similar, but both
jet sources have been defined as point sources. Pixel size is 0$\farcs$492.}
\label{sim} 
\end{figure}

To characterize the jet morphology further, we smoothed the image using CIAO task
aconvolve, treating the jets and the stellar PSF separately.
In Fig.~\ref{jetim}b we used a Gaussian sigma of 0.75 pixels
for the area outside the star, and 0.5 pixels for the stellar image. The smoothed
image clearly shows the extended morphology of the two jet sources, reminiscent of 
the optical image at least for the SW  jet \citep{dougados00}. 

Figure~\ref{sim} shows simulations of two different 
jet source shapes, based on the MARX software. 
In the left figure, the SW and NE jets are composed of a Gaussian with
sigma = $1\arcsec$, and the SW jet additionally contains a  1-D linear source of 
$2\arcsec$ length, positioned radially outside the stellar PSF at a position angle of 225~deg. 
The linear feature contains $\approx 1.4$ times as many counts as the Gaussian source. 
The right figure shows a simulation in which both jet sources are point sources.
\footnote{The half-energy 
radius of the point response function of an ACIS-S on-axis source is
about 0$\farcs$41 or $\approx$ 1 pixel, and the 80\% encircled-power radius is $\approx 0\farcs 7$
or nearly 2 pixels (see {\it Chandra} Proposers' Observatory Guide v.9). We also simulated the 
PSF in the {\it Chandra Ray Tracer} (ChART; \citealt{carter03}) 
software (http://cxc.harvard.edu/soft/ChaRT/cgi-bin/www-saosac.cgi)
for the position of DG Tau and the boresight parameters of ObsID 4487 and 7246. The PSF
was found to be compact, its core being very slightly elongated along a SE-NW axis, as can
also be seen in Figs.~\ref{jetim} and \ref{sim}. There is no extension in the directions
of the jets (SW-NE).}

The total number of counts
in each simulated jet feature is similar to the numbers in the real 90~ks observation.
The observation resembles the simulation of extended sources;
it is not compatible with one point source per jet. A more quantitative comparison
will need longer exposures of these features.

We next generated ``hardness'' images, using two complementary procedures.
In the first case, each individual {\it count} was attributed a color as a function of its energy.
Energies in the 0.55--0.7~keV range are represented in red, energies 
in the 1.5--1.75~keV range in blue. In the 0.7--1.5~keV range, colors
change continuously from red to yellow to blue. The resulting image was smoothed, using a
Gaussian with sigma = 0.85 pixels - see Fig.~\ref{jetim}c.

For an alternative representation, we extracted three {\it pixel} images smoothed as 
before (Gaussian smoothing  with sigma = 0.75 pixels),  a ``red'', ``green'',
and ``blue'' image for the 0.55--0.85~keV, 0.85--1.2~keV, and 1.2--1.75~keV ranges, 
respectively. The combined false-color image is shown in Fig.~\ref{jetim}d.

Both hardness images show that the counter jet is harder, with photon energies mostly above
1~keV, while the forward jet shows a mixture of softer and harder counts.

\begin{table}[t!]
\caption{Results from spectral fit to jet sources. Errors give $1\sigma$ confidence ranges, and EM and $L_{\rm X}$
         values are for one jet each}
\label{jet}
\begin{tabular}{lcc}
\hline
\hline
                                          & SW Jet:                             & NE Jet:            \\
\hline
 $N_{\rm H}^{a,b}$                        &  $3.0^{+3.0}_{-3.0}\times 10^{21}$~cm$^{-2}$ & 5.7$\times 10^{21}$~cm$^{-2}$  	 \\ 
 $\Delta N_{\rm H}$                       &   \multicolumn{2}{c}{2.7$^{+1.5}_{-1.1}\times 10^{21}$~cm$^{-2}$}         	          \\ 
 $T^b$   		                  &   \multicolumn{2}{c}{$3.4^{+5.2}_{-1.1}\times 10^6$~K}                     \\ 
 EM$^b$                                   &   \multicolumn{2}{c}{$1.0^{+6.3}_{-0.8}\times 10^{51}$~cm$^{-3}$} 	 \\ 
 $L_{\rm X}$ [0.1-10~keV]                 &   \multicolumn{2}{c}{$1.2\times 10^{28}$~erg~s$^{-1}$} 	                \\ 
 \hline
\end{tabular}
\begin{list}{}{}
\item[$^{\mathrm{a}}$]{The two $N_{\rm H}$ values are coupled through the fit parameter  $\Delta N_{\rm H}$. 
                       Therefore, no independent errors are given for the NE jet.}
\item[$^{\mathrm{a}}$]{$N_{\rm H, forward}$ and EM are, within $1\sigma$, poorly constrained to low values and $T$ is poorly 
                      constrained to high values because there is a second (probably unphysical) minimum in the statistic,
                      with $N_{\rm H, forward} \approx 0$, $T = 6.8$~MK, and very small EM. See text for details.}
\end{list}
\end{table}

\subsection{Spectral analysis of the jet sources}

\begin{figure}[t!]
\includegraphics[angle=-90,width=8.3cm]{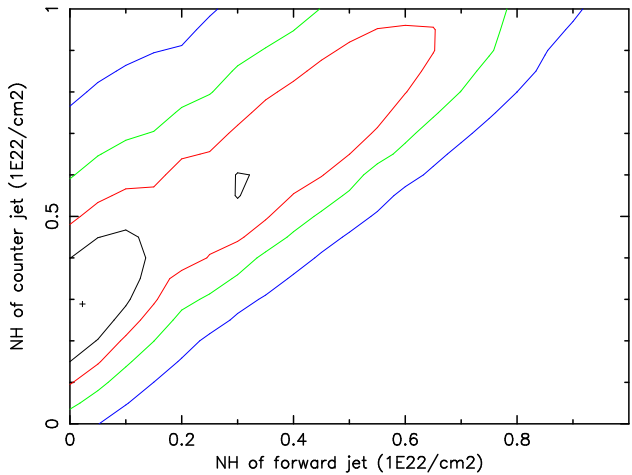}
\caption{Confidence regions for $N_{\rm H}$ of the counter jet versus $N_{\rm H}$ 
of the forward jet, using an isothermal plasma component per jet (with equal EMs and temperatures).
The contours show the $1\sigma$ (68\%, in red), 90\%, (in green), and 99\% (in blue) confidence regions for two
interesting parameters based on the C statistic. The black contour is plotted to outline the regions of
lowest C statistic (referring to $\Delta$C = 1). The cross marks the location of the lowest C statistic, although the minimum  
at $N_{\rm H, forward} = 0.3$ and $N_{\rm H, counter} = 0.57$ is more likely to represent a physical solution.  
}
\label{contours} 
\end{figure}

Only simplistic models can be fitted to the spectra of the jets, given the small number of
counts in the two jet sources.
We experimented with isothermal plasma components (using the vapec model in XSPEC, with the abundances
mentioned above) subject to  photoelectric absorption (using the wabs model in XSPEC). Some
plausible assumptions were made to keep the number of fit parameters low: Because the jet morphologies
and the lengths of the two jets are similar, we adopted equal emission measures and (for similar heating
mechanisms, e.g., shocks with similar velocity) equal temperatures. The absorption column densities, 
$N_{\rm H}$, remained different for each source because different gas components (e.g., the star's extended
circumstellar disk) may absorb the jets differently, as also suggested by the hardness images discussed
above. The initial fit parameters were therefore the emission measure (per jet), a single electron temperature (for both jets), and
two $N_{\rm H}$ values. We found, however, that the absolute values of $N_{\rm H}$ are poorly constrained due
to numerical cross-talk with the temperature
(lower $N_{\rm H}$ can be compensated by higher temperatures), while the {\it difference} 
$\Delta N_{\rm H} = N_{\rm H, counter}-N_{\rm H, forward}$ between the absorption of the 
the counter jet and the forward jet
is sufficiently well confined. This is illustrated in Fig.~\ref{contours} where the confidence regions are 
shown on the $N_{\rm H, counter}$ vs. $N_{\rm H, forward}$ plane.
The region of optimum $N_{\rm H}$ values shows a strong and nearly 
linear correlation between the two values, in such a way that their difference is well constrained. 
We have therefore chosen to fit $N_{\rm H, forward}$ and $\Delta N_{\rm H}$, while the 
value of $N_{\rm H, counter}$ can be derived from any pair of these two parameters.

The best-fit $\Delta N_{\rm H}$ is $2.7\times 10^{21}$~cm$^{-2}$, with a $1\sigma$ range  
of $(1.6 - 4.2)\times 10^{21}$~cm$^{-2}$ and a 90\% range  of $(0.9 - 5.2)\times 10^{21}$~cm$^{-2}$.
A differential absorption column between the two jets is thus very likely, and is in fact expected as
discussed below.  The plot shows two minima, the slightly deeper one located at $N_{\rm H, forward} 
\approx 0.02\times 10^{22}$~cm$^{-2}$ and $N_{\rm H, counter} \approx 0.29\times 10^{22}$~cm$^{-2}$, 
and a slightly shallower one at $N_{\rm H, forward} 
\approx 0.30\times 10^{22}$~cm$^{-2}$ and $N_{\rm H, counter} \approx 0.57\times 10^{22}$~cm$^{-2}$. 
The two solutions agree at the 1$\sigma$ level; however, the
former solution requires a high best-fit temperature (0.59~keV) which, together with an $N_{\rm H}$
value that is unusually low for Taurus pre-main sequence stars \citep{guedel07a}, suggests that the solution is
unphysical. We adopt the second solution  as the more reasonable ``best-fit'' in the following although we
repeat that only $\Delta N_{\rm H}$ can be sufficiently well constrained.

Despite the large errors of the fit parameters, three features are noteworthy. First, the absorption column density
toward the forward (SW) jet is small ($\approx 3\times 10^{21}$~cm$^{-2}$), 
agreeing with $N_{\rm H}$ of the soft but not the hard stellar component (see below). 
Second, the absorption toward the counter jet is  higher, compatible with the increased
hardness discussed in Sect.~\ref{morph}. And third, the electron temperature of the jet 
sources is low ($\approx$3.4~MK) compared with coronal temperatures of T Tauri stars 
(e.g., \citealt{telleschi07b}).

We do not give errors for $L_{\rm X}$ as these depend very sensitively but non-trivially
on the rather uncertain absorption column densities (see \citealt{guedel07a} for estimates
of lower limits to the uncertainties of $L_{\rm X}$ of many X-ray sources in the Taurus
star-forming region). Based on the best-fit values, the  total X-ray
output from the jets outside the stellar PSF (radius of $1\farcs 4-1\farcs 7$) would amount to 
a few times $10^{28}$~erg~s$^{-1}$ or 10--20\% of the {\it stellar} soft component (see below).

\begin{figure}[t!]
\includegraphics[angle=-90,width=8.8cm]{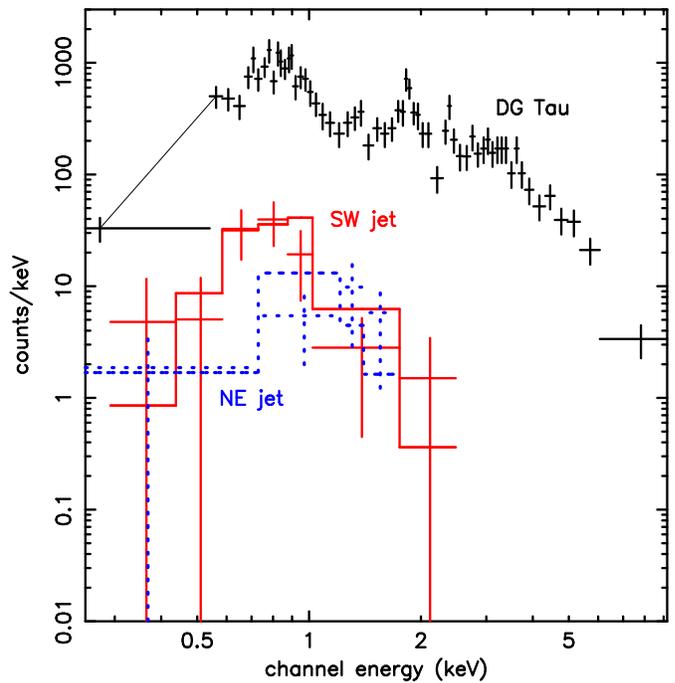}
\caption{Co-added ACIS-S spectra of the DG Tau point source (upper spectrum, black; the lowest-energy 
         bins are connected for clarity) and the
         two jets (lower spectra; spectrum of less absorbed forward jet is shown solid, in
	 red, while spectrum of the more strongly absorbed counter jet is shown dotted, in blue).
	 For the jet sources, the solid histogram shows the joint spectral fit. The spectrum of DG Tau
	 has been binned such that at least 15 counts are contained in each bin; the jet spectra have
	 been binned arbitrarily to produce non-zero bins from $\approx 0.3$~keV to $\approx 2$~keV.
	 }
\label{specfigure} 
\end{figure}

\begin{table*}[t!]
\caption{Results from spectral fits for the DG Tau point source. Errors give 90\% confidence limits}
\label{spec}
\begin{tabular}{lllllll}
\hline
\hline
 \noalign{\smallskip}
 & \multicolumn{5}{c}{\it Chandra:}	     & {\it XMM-Newton:}     \\
 & \multicolumn{5}{c}{\hrulefill}	     & \hrulefill           \\
 \noalign{\smallskip}
 ObsID                                    &  4487                  &  6409		     & 7247		     & 7246		     & all {\it Chandra}    & 0203540201$^a$	 \\
 Epoch                                    &  Jan. 2004             &  Dec. 2005 	     & Dec. 2005	     & Apr. 2006	     & 2004-2006	    & Aug. 2004 	    \\
 counts [0.2-7~keV]                       &  388                   &  204		     & 142		     & 349		     & --		    & 634		    \\
 count rate [ct~s$^{-1}$, 0.2-7~keV]$^b$  & 0.0131(7)              &  0.0126(9)              & 0.0089(7)             & 0.0125(7)             & --                   & 0.0206(10)          \\
 {\bf Soft component:}                    &                        &			     &  		     &  		     &  	            & 		     \\
 count rate [ct~s$^{-1}$, 0.2-1.1~keV]$^b$& 0.0046(4)              &  0.0041(5)              & 0.0041(5)             & 0.0046(4)             & --                   & 0.0083(6)          \\
 $N_{\rm H, 1}$  [$10^{21}$~cm$^{-2}$]    &  1.4$^{+1.8}_{-1.0}$   &  0.95$^{+2.3}_{-0.93}$  & 4.5$^{+4.1}_{-1.9}$   & 2.0$^{+2.6}_{-1.7}$   & 1.3$^{+1.1}_{-0.6}$  & 1.1$^{+1.4}_{-0.8}$		   \\ 
 $T_1$ [MK]  		                  &  4.6$^{+2.3}_{-1.3}$   &  6.0$^{+1.7}_{-2.7}$    & 3.2$^{+1.5}_{-1.5}$   & 3.6$^{+2.3}_{-1.0}$   & 4.5$^{+0.7}_{-0.9}$  & 3.7$^{+0.9}_{-0.5}$	   \\ 
 $L_{\rm X, 1}$ [erg~s$^{-1}$, 0.1-10~keV]&  $1.1\times 10^{29}$   &  $8.2\times 10^{28}$    & $4.5\times 10^{29}$   & $1.7\times 10^{29}$   & $1.0\times 10^{29}$  & $0.96\times 10^{29}$	   \\ 
 {\bf Hard component:}                    &                        &		             &		             &		             &		            & 		      \\
 count rate [ct~s$^{-1}$, 1.7-7~keV]$^b$  & 0.0063(5)              &  0.0068(6)              & 0.0031(4)             & 0.0066(5)             & --                   & 0.0099(7)          \\
 $N_{\rm H, 2}$ [$10^{21}$~cm$^{-2}$]     &  21$^{+8}_{-6}$ 	   &  32$^{+16}_{-11}$       & 15$^{+14}_{-10}$      & 33$^{+11}_{-12}$      & --		    & 18$^{+6}_{-6}$  \\ 
 $T_2$ [MK]  		                  &  31$^{+21}_{-9}$       &  27$^{+25}_{-10}$       & 34$^{+77}_{-16}$      & 23$^{+23}_{-6}$       & --		    & 69$^{+119}_{-25}$       \\ 
 $L_{\rm X, 2}$ [erg~s$^{-1}$, 0.1-10~keV]&  $6.9\times 10^{29}$   &  $1.0\times 10^{30}$    & $2.8\times 10^{29}$   & $1.1\times 10^{30}$   & --		    & $9.6\times 10^{29}$     \\ 
 \hline
\end{tabular}
\begin{list}{}{}
\item[$^{\mathrm{a}}$] Average properties for the hard component during {\it XMM-Newton} EPIC PN observations (including flares); counts for 24932~s of low-background data  
\item[$^{\mathrm{b}}$] parentheses give errors in last digit
\end{list}
\end{table*}

\section{The DG Tau X-ray source}

The two-component spectral phenomenology of TAX sources \citep{guedel07b} 
is present in each of the new {\it Chandra} 
observations. The spectral-fit results are reported in Table~\ref{spec}, together with the results from the 
{\it XMM-Newton} observation taken from \citet{guedel07b}. The rather short exposures
obtained in 2005-2006 led to considerable uncertainties in the derived parameters.
This is particularly evident in the absorption column density, $N_{\rm H, 1}$, the 
electron temperature, $kT_1$, and consequently the X-ray luminosity, $L_{\rm X, 1}$, 
of the softer spectral component for the shortest observation,
ObsID 7247, comprising only 142~cts. On the other hand, the count rates of the soft component
agree within the error bars for all {\it Chandra} observations (Table~\ref{spec}), and no 
variability was  recorded in the soft component during the individual exposures. Because the 90\% error 
bars of $N_{\rm H, 1}$ and $kT_1$ also strongly overlap for the four {\it Chandra} exposures, 
we performed a joint fit of these spectra in the energy range of 
0.2-1.1~keV, reported in the penultimate column in the table. 
The resulting parameters of the soft component compare very favorably
with the {\it XMM-Newton} EPIC PN results (last column).

Because the hard component is 
variable (Table~\ref{spec}; slow, non-periodic modulations by a factor of $\approx$2 were present 
in ObsID 7246 although larger flares were absent), 
agreement between the fit parameters is not expected, but we notice that
the hydrogen absorption column density,  $N_{\rm H, 2}$, agrees in all observations within the 90\%
error ranges. It is remarkable that  $N_{\rm H, 2}$ is approximately 20 times higher than
$N_{\rm H, 1}$.

For illustration purposes, we show the combined {\it Chandra} spectrum in Fig.~\ref{specfigure} but 
again emphasize that spectral fits were performed for the individual, unbinned spectra (illustrations of
spectral fits to binned data are shown in \citealt{guedel05} for {\it Chandra} and in \citealt{guedel07b} 
for {\it XMM-Newton}).

\section{Discussion}

Our {\it Chandra} observations add unprecedented information to the X-ray emission model of DG Tau. 
In particular, we report the detection of an extended, bipolar X-ray jet down to the stellar
PSF and out to a distance of about 5\arcsec. This is the first double-sided X-ray jet reported
from a pre-main sequence star.

\begin{figure}[t!]
\hskip -0.2truecm\includegraphics[angle=0,width=9.0cm]{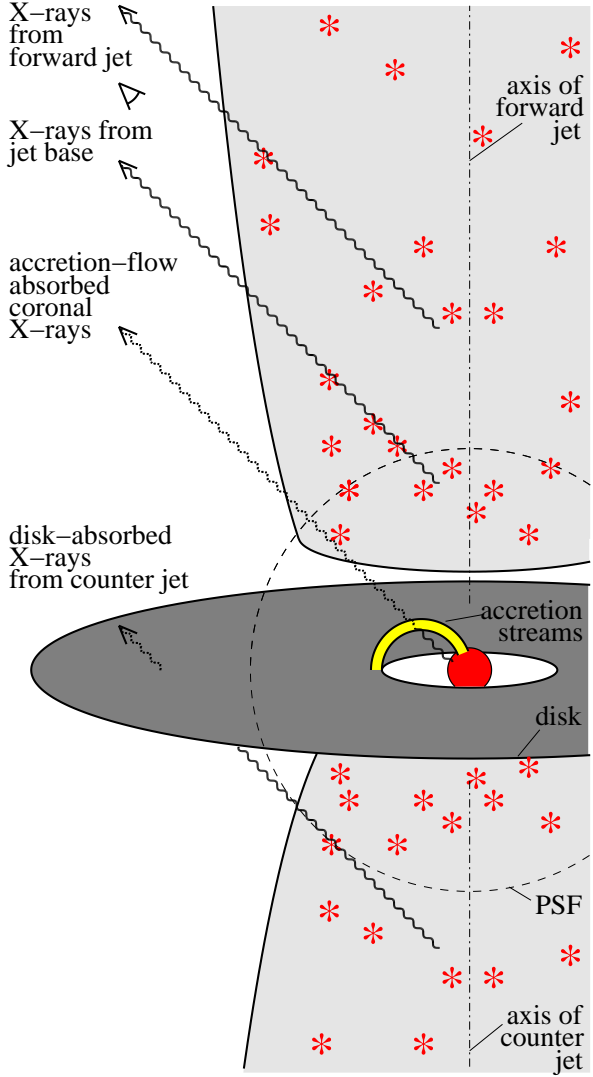}
\caption{Sketch of the proposed model of the DG Tau environment. Features are not drawn to scale. The observer
is located toward the upper left. Four X-ray source regions are shown in red (the stellar coronal source, and
asterisks symbolically marking emission regions in the jet) and their absorbing media are schematically shown. 
X-rays seen by the observer are sketched by the wavy lines, described on the left, from top to bottom: 
X-rays from the weakly absorbed, spatially resolved 
X-ray source of the forward jet; from the 
spatially unresolved but spectrally identified soft X-ray source closer to the base of 
the jet; from the hard, stellar coronal emission absorbed by the infalling accretion streams;
and from the  spatially resolved X-ray emission from the counter jet, slightly absorbed by
the intervening disk gas. The dashed circle denotes the observing PSF of the star; features inside the
PSF are unresolved from the stellar X-ray source.}
\label{model} 
\end{figure}

The ACIS-S images reveal the presence of jets in the SW and NE directions, coincident with
the direction of the optical jets. The NE jets appears to be slightly harder and also shows fewer
photons in a similar extraction region. Further, the spectra of the DG Tau point source reveal
a TAX spectral morphology. The DG Tau system thus hosts at least four
distinct X-ray sources of different origin and subject to different gas absorption columns, as
illustrated in the sketch shown in Fig.~\ref{model}. These components are:
\begin{enumerate}
\item[1.] a weakly absorbed, diffuse, soft component along the forward-jet axis;\\
\item[2.] an more strongly absorbed, diffuse, soft component along the counter-jet axis;\\
\item[3.] a weakly absorbed, compact, non-variable, soft component (within the stellar PSF);\\
\item[4.] a strongly absorbed, compact, flaring, hard component (within the stellar PSF).\\
\end{enumerate}
We now discuss the various features seen in the X-ray spectra and images, and propose a model
consistent with all observed features. 

\subsection{The X-ray jets: Physical properties}

In this section, we will be
interested as to how the jet cools, and why no X-ray emission is seen at distances beyond 5\arcsec\ from
the star. There are further prominent bow shocks outside
the regions discussed here, in particular a structure that was located at 10\farcs 7 from DG Tau in observations
obtained in 1986 December \citep{eisloeffel98} and still within $a\approx 12$\arcsec\ a decade later 
\citep{stapelfeldt97}. Using a proper motion of 0\farcs 15--0\farcs 3 per year \citep{eisloeffel98, dougados00},
this bow shock should now have expanded to about 15\arcsec. Obviously, the DG Tau jet consists of a series of
knots that are frequently ejected from the star \citep{eisloeffel98}. If the more distant knots were similar to those
now seen in X-rays when at similar distances from the star, then the heated gas must have cooled, or the emission
measure have decreased below our detection limit because we have found no significant X-ray source at those
distances. Given  the time to travel from $\approx 5$\arcsec\ to $\approx 11$\arcsec (the region behind the
outer bow shock, in the ``X-ray free'' lower right corner of Fig.~\ref{jetim}), we adopt an X-ray decay time (with respect 
to the range of sensitivity of the X-ray detectors) of $\approx 20-40$~yrs. 

To explore the potential cooling mechanism(s), we first discuss two extreme cases of cooling: 
cooling by expansion, i.e., adiabatic cooling; and radiative cooling without expansion. 
We will then discuss cooling if contributions from both cooling mechanisms contribute.

\subsubsection{Cooling by expansion}\label{expansion}

The DG Tau jet significantly expands with increasing distance from the star. Transverse expansion has been
measured; various  features within a few arcsec of the star show an unusually large opening angle in the range of 
$\beta \approx 11-27$~degrees (based on full width at half maximum, after \citealt{dougados00}, see also 
\citealt{eisloeffel98}). In contrast, radial expansion (i.e., stretching of volumes in radial direction, along 
the linear flow)  is unlikely to be important as this would require strong jet acceleration at distances of 
several arcseconds, or selective deceleration of regions closer to the star, for which no evidence has been 
reported.

There are two consequences of transverse expansion: a decrease of the EM and adiabatic cooling (in the
absence of radiation), both leading to a decay of the X-ray emission in time. 

The cross-section area of the jet at distance $r$ from the star is
\begin{equation}\label{cross}
A(r) = \pi r^2\tan^2{\beta\over 2}.
\end{equation}
or for small changes in $A(r)$,
\begin{equation} 
{dA\over A} = 2{dr\over r}.
\end{equation} 
For the time span of 2 years, 
$dr = 0\farcs 3$--$0\farcs 6$, and therefore $dA/A \approx $0.12--0.24
at a distance of 5\arcsec. The electron density decreases by the same factor as the volume increases, and 
therefore the emission measure, EM $= n_{\rm e}^2V$, decreases by 12--24\%  at a distance of $\approx$5\arcsec\ 
within 2 years. Such a decrease is too small to be significantly measured in our 2004--2006 data. 

However, for a distance of, say, 8\arcsec--11\arcsec\ (the lower-right area in Fig.\ref{jetim}), 
$A(r)/A(r_0) = \left(r/r_0\right)^2 \approx$ 2.6--4.8.
The emission measure for a given mass element in the jet thus  decreases by a factor of 2.6--4.8 from a distance of 
5\arcsec\ to 8--11\arcsec\ from the star due to expansion alone. Because the gas is optically thin to X-ray emission, the
X-ray surface brightness decreases by a factor of 2.6$^2$--4.8$^2$ = 6.8--23, making its detection at this distance
against the background level very difficult in our exposures.

At the same time, the plasma cools by expansion. In the limiting case, we neglect radiative losses, i.e.,
the gas cools adiabatically. Then, 
\begin{equation}\label{adiabat}
TV^{\gamma-1} = {\rm constant}
\end{equation} 
with $\gamma = 5/3$ for monatomic gas, and for small changes in $T$,
\begin{equation}\label{adiabat2}
{dT\over T} = (1-\gamma) {dV\over V} = -{2\over 3}{dA\over A} = -{4\over 3}{dr\over r}.
\end{equation}
Using $dA/A \approx 0.12-0.24$ for a two-year interval, we find $|dT/T| \approx 0.08-0.16$. Again, 
this temperature decrease is too small to be measured during our observing interval. 

For long intervals, however, 
\begin{equation}
{T(r)\over T(r_0)} = \left({A(r)\over A(r_0)}\right)^{-2/3}  = \left( {r\over r_0}\right)^{-4/3}
\end{equation}
which, for $r= $8\arcsec--11\arcsec, the above projected jet velocities and $r_0 = 5$\arcsec\ 
yields $T(r)/T(r_0) \approx$ 0.35--0.53, i.e. $T(r) \approx $1.2--1.8~MK. 

A temperature decrease will also affect the count statistics in the detector, because the effective
area decreases toward the softer energy range. We simulated the spectrum of the forward jet in XSPEC, adopting a much 
higher flux for better statistics and further changing  its temperature. We then measured the simulated 
count rate in  the 0.6--1.7~keV range. Lowering $kT$ to 0.249--0.272~keV ($T$ = 2.9--3.2~MK), which corresponds to 
$|dT/T| \approx$ 0.08--0.16, reduces the count rate by 11--26\%. Together with the count rate drop from
the decrease of the EM (see above), the total drop of the detected count rate would thus be $\approx$ 23--50\%. 
It is possible that this decrease affected our measured count rates,
but we cannot prove such cooling effects as the count-rate decrease was not statistically significant during
the two years covered by our observations. Furthermore, as some of the material in the detected jet
structure may cool, new hot material is likely to be replenished from closer to the star, as we have
found the X-ray structure to be extended toward the stellar PSF.

On the other hand, considering a longer time span of 20 years, i.e., and expansion from 5\arcsec\ to 8\arcsec--11\arcsec, 
therefore lowering the plasma temperature to
1.8~MK, the count rate decreases by 83\%. This, together with the strong decrease of the EM, reinforces
our finding that a cooling, expanding X-ray jet starting under presently observed conditions at 5\arcsec\ distance 
will be difficult to detect at distances of 8--11\arcsec, i.e., the region just behind a bow shock 
described by \citet{eisloeffel98}.

In summary, adiabatic expansion alone should lead to the disappearance of the X-ray jet at distances significantly 
beyond 5\arcsec\ in {\it Chandra} exposures such as ours. Our exposures are not sufficient, on the other
hand, to verify a cooling effect during the two-year time span. 

\subsubsection{Cooling by radiation}\label{radiation}

Radiative cooling may be significant as well.  We cannot estimate the radiative cooling
time because the plasma density is unknown, or in other words, it is not clear whether the observed
X-ray radiation extracts a significant fraction of the thermal energy of the hot plasma within the 2--20~yrs of 
interest here.  We will, on the other hand, study consequences for the X-ray jet {\it if}
radiative cooling is the dominant loss mechanism. The volume of the X-ray detected source of the
forward jet is approximately (see Sect.~\ref{morph}) 
\begin{equation}
V = {3\arcsec * (0\farcs 5)^2\pi\over \sin(38~{\rm deg})} * \left(2.1\times 10^{15}\right)^3~{\rm cm}^{3},
\end{equation} 
where 38~deg is the jet inclination \citep{eisloeffel98}, we have adopted a length outside the 
PSF and a cross-sectional radius of the X-ray jet of 3\arcsec\ and 0\farcs 5, respectively 
(Sect.~\ref{morph}), and the final constant gives the conversion from arcseconds to cm at the 
distance of Taurus. We thus find $V = 3.6\times 10^{46}$~cm$^{-3}$ but note that the X-ray gas
may occupy only a fraction of this volume, defined by the volume filling factor $f$. 

With the best-fit EM from Table~\ref{jet}, the density of the X-ray emitting jet gas is 
\begin{equation}\label{density}
n_{\rm e} = \left({{\rm EM}\over fV}\right)^{1/2} \approx  {170\over f^{1/2}}~{\rm [cm}^{-3}].   
\end{equation}
The thermal energy decay time then is, assuming that energy is lost by radiative cooling (i.e., decrease of $T$) only, 
\begin{equation}\label{radloss}
\tau = {3kT\over n_{\rm e}\Lambda(T)} \approx 1.4\times 10^5 T f^{1/2}~ {\rm [s]} \approx 15400 f^{1/2}~ {\rm [yr]}
\end{equation}
where $\Lambda(T) \approx 1.7\times 10^{-23}$~erg~cm$^3$s$^{-1}$ is the cooling function (evaluated at 3.4~MK for our
model, derived from apec for the 0.001-100~keV interval), and in the last equation we have adopted the X-ray 
measured jet temperature of 3.4~MK. 

Here, the energy decay reflects exclusively in a decrease of the temperature, $T$, while the emission measure
remains constant. As before, we model the count rate reduction for decreasing temperatures for the ACIS-S
detector. Given that we detected 18 cts in the forward jet, we now seek the temperature for which the same
emission measure results in a detection limit of only 5--6 detected counts, i.e., a count rate reduced to 
25--30\%. For fewer counts,
the source will be difficult to detect, depending also on the spatial distribution of the counts (we recall
that approximately one count will be due to contamination from background radiation and the stellar PSF).
Using the detector response, we found that a count rate reduction to 25--30\% corresponds to a temperature
reduction to 50--61\% of the initial value, i.e., a time interval of 0.49--0.56 e-folding decay times of the energy 
(Eq.~\ref{radloss}).

If the jet should become undetectable in an observation like ours after 20--40~yr,
i.e., for an energy e-folding decay time of 10--22~yr, we require 
$f = 4\times 10^{-7}$--$2\times 10^{-6}$, or densities of $(1.2-2.6)\times 10^5$~cm$^{-3}$.

\subsubsection{Cooling by expansion and radiation}

If radiative losses are significant during the time span of interest here, then the adiabatic approximation
in Sect.~\ref{expansion} breaks down. On the other hand, if we accept that the hot plasma is subject
to expansion as observed for the cooler optical jet, then the treatment of radiative losses in Sect.~\ref{radiation}
is also not sufficient. Both cooling terms must be combined. 

We consider
 the first law of thermodynamics,
\begin{equation}\label{law}
dU + \delta W = \delta Q.
\end{equation}
Here, $dU$ is the change of the internal energy, $U = 2\alpha N_{\rm e}kT$, where $N_{\rm e}$ is the total number of electrons
and $\alpha = 1/(\gamma -1) = 3/2$ for monatomic gas (for simplicity, we adopt a hydrogen plasma with a total number
of particles $N = N_{\rm p} + N_{\rm e} = 2N_{\rm e}$; the difference to a realistic plasma is not significant, 
the electron density in the latter being only about 10\% larger than the ion density). Further, $\delta W = pdV$ is 
the pressure work done by the gas, with a pressure of $p = 2n_{\rm e}kT$. If radiative losses are present, we set
\begin{equation}
\delta Q = - n_{\rm e}(t)^2V(t)\Lambda(T)  dt
\end{equation}
where $\Lambda(T)$ is the temperature-dependent cooling function (in erg~cm$^{3}$~s$^{-1}$) and $n_{\rm e}$ is 
the electron density. Substituting these expressions into Eq.~(\ref{law}), we find
\begin{equation}\label{energy}
\alpha {dT\over T(t)} + {dV\over V(t)} = -{n_{\rm e}(t)\Lambda(T)\over 2kT(t)}dt.
\end{equation}
Note that in the absence of radiative losses, Eq.~(\ref{energy}) is equivalent to equations~\ref{adiabat} and \ref{adiabat2}
for an adiabatic process. Radiation makes the problem explicitly time-dependent; the initial electron 
density, $n_{\rm 0} = n_{\rm e}(t=0)$,  is introduced as a free parameter.

\begin{figure}[t!]
\hskip -0.2truecm\includegraphics[angle=0,width=9.0cm]{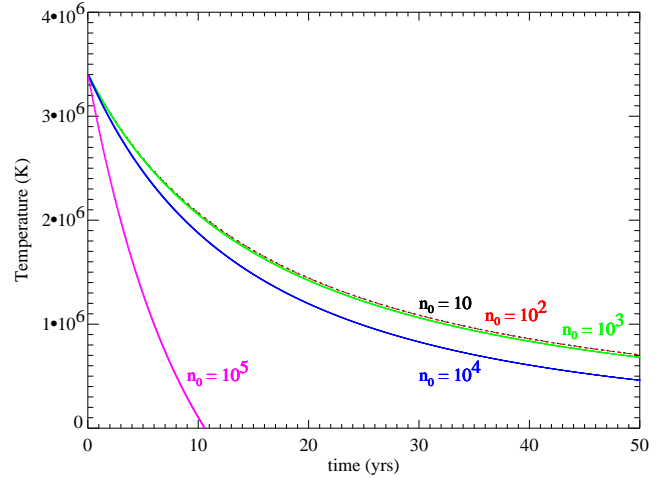}
\caption{Cooling behavior of the jet plasma, assuming transverse expansion as observed in the optical jet, and
         radiative losses for various initial electron densities, $n_0 = n_{\rm e}(t=0)$ at a stellar distance 
	 of $r_0 = 5$\arcsec. The curves are color-coded for initial densities given in the plot. The two curves for
	 $n_0 = 10$~cm$^{-3}$ and $n_0 = 100$~cm$^{-3}$ are nearly coincident.}
\label{cool} 
\end{figure}

We define the observed expansion of a given volume element by the radial distance along the jet axis, 
$r(t) = r_0 + vt$. Then, $V \propto r^2$ (in the absence of expansion along the jet axis), 
and $n_{\rm e} \propto V^{-1} \propto r^{-2}$. Substituting these expressions, we find 
\begin{equation}
\alpha {dT\over T(t)} = -\left[  {2v\over r(t)} + \left({r_0\over r(t)}\right)^2 {n_0\Lambda(T)\over 2kT(t)} \right] dt.
\end{equation}
We have determined the run of $T$ numerically, starting with $T_0 = 3.4$~MK and with various $n_0$ values, taken
to be the initial electron density at $r_0$ = 5\arcsec\ (not yet considering a strict lower limit to $n_0$ - see below). 
The cooling function was approximated by $\Lambda(T) = 4.83\times 10^{-23} - 9.75\times 10^{-24}~T$~erg~cm$^{3}$~s$^{-1}$ 
($T$ in MK) for the range of $T = $1--3.5~MK,
as found from an integration of the spectral energy distribution for an isothermal plasma in XSPEC, using the
vapec code with our adopted element abundances (Sect.~\ref{obs}). For $T < 1$~MK, we kept $\Lambda$ constant at 
$4\times 10^{-23}$~erg~cm$^{3}$~s$^{-1}$.  We used an average projected jet velocity of 0.225\arcsec~yr$^{-1}$.
The results are shown in Fig.~\ref{cool}.
We see that - in agreement with the estimates for pure radiative losses - the cooling process
is dominated by expansion up to densities of several times $10^3$~cm$^{-3}$, i.e., cooling is
adiabatic. Only for densities $\ga 10^4$~cm$^{-3}$ does radiation matter on time scales of tens of years.

\subsubsection{Implications from jet cooling}

Because the electron density in the jet X-ray source is unknown, the importance of radiative cooling cannot be directly
assessed. We will, however, discuss the limiting case of low densities. Equation~(\ref{density}) gives the lowest density
for the maximum volume filling factor, $f = 1$, namely $n_{\rm e} = 170$~cm$^{-3}$.  This implies an upper limit for
the thermal energy of the hot gas of $3n_{\rm e}kTV \approx 8.6\times 10^{39}$~erg in the forward jet alone or 
$\approx 1.7\times 10^{40}$~erg in both jets together.

The mass outflow rate in the DG Tau jets is $6\times 10^{-8}~M_{\odot}~$yr$^{-1}$ (twice the value given
by \citealt{lavalley00} for the forward jet)
with a characteristic flow velocity of 300~km~s$^{-1}$ (e.g., \citealt{eisloeffel98, dougados00}).
The kinetic energy rate is therefore $\dot{M}_{\rm jet}v_{\rm jet}^2/2 \approx 1.7\times 10^{33}$~erg~s$^{-1}$. 
The total jet X-ray luminosity is, according to Table~\ref{jet}, $2.4\times 10^{28}$~erg~s$^{-1}$, i.e., a 
fraction of $1.4\times 10^{-5}$ of the kinetic energy dissipates in X-rays after transformation to thermal 
energy of the hot plasma.

The jet out to a distance of 5\arcsec\ forms within $\approx 20$~yrs based on the measured proper motion.
The mass outflow rate then indicates that $\approx 1.2\times 10^{-6}~M_{\odot}$ are contained in
the jets, with a total kinetic energy of $1.1\times 10^{42}$~erg. Therefore, a maximum of 1.6\% of the kinetic 
energy is transformed into thermal energy in the hot plasma.

We can now investigate whether the pressure in the hot gas contributes to the expansion of the jet.
The bulk gas observed in optical lines has a density of $10^3$~cm$^{-3} -10^4$~cm$^{-3}$ at distances of  
a few arcsec from the star, with an ionization fraction approaching unity \citep{lavalley00}. 
The temperature is of order $10^4$~K \citep{hamann94}. The gas pressure is therefore of order 
$P_{\rm bulk} \approx (3\times 10^{-9}-3\times 10^{-8})$~dyn~cm$^{-2}$. For the X-ray emitting gas, we estimate, 
from the above values, a {\it minimum} pressure for $f = 1$, namely $P_{\rm hot} \approx (1.6\times 10^{-7})$~dyn~cm$^{-2}$. 
A filling factor of unity is impossible because only a small fraction of the energy heats the hot plasma while
cooler gas predominates. It thus appears that the hot gas pressure is grossly out of equilibrium with its environment,
{\it contributing to the transverse jet expansion discussed in this section.}

We emphasize that this argumentation holds regardless of the exact filling factor or of the relevance of 
radiative cooling. It is a minimum-density estimate related to the energy decay in Eq.~(\ref{radloss}).

\subsection{Heating the X-ray jets}

How are the jets heated to the observed temperatures? Most of the conventionally detected emission
from jets of pre-main sequence stars comes from low-ionization transitions such as [O\,{\sc i}], [N\,{\sc ii}], or 
[S\,{\sc ii}],
indicating temperatures of no more than a few thousand K.  Hotter gas has been identified, e.g.,
in emission from He\,{\sc i}, [O\,{\sc iii}], C\,{\sc iii}, and O\,{\sc iv}, giving evidence for hot winds \citep{cohen85, takami02, 
dupree05} although some of these claims have been questioned, and maximum wind temperatures of 
only $2\times 10^4$~K have been derived \citep{johns07, kwan07}. 
Measured shock velocities in the jets of DG Tau are of order 50--100~km~s$^{-1}$ \citep{lavalley00},
sufficient to heat gas to a few $10^5$~K but not to the observed 3--5~MK.

A possibility is that a fraction of the jet gas collides directly with the interstellar medium, in which case
the shock velocity would be several 100~km~s$^{-1}$ and therefore in principle sufficient to heat a fraction
of the gas to X-ray emitting temperatures. Alternatively, magnetic heating may be in operation. Magnetic
fields are commonly invoked to explain jet acceleration near the star and the disk, and also 
to collimate jets during their propagation \citep{blandford82, uchida85, koenigl00, shu00}.

Ambipolar diffusion heating results from the separation of charged particles from neutrals across 
magnetic fields. The process has been described in detail by \citet{safier92}, \citet{garcia01a}, and 
\citet{garcia01b}.  It is most efficient as long 
as the gas is weakly ionized and therefore not relevant for our hot plasma. Self-consistent jet 
models converge to jet temperatures of $\approx 10^4$~K due to ambipolar diffusion heating.

Ohmic dissipation of currents  operates in highly ionized gas. Initially jet-aligned magnetic 
fields are wound up due to rotation, producing helical fields which drive currents. Depending
on $\nabla\times \mathbf{B}$, sufficient dissipation may be achieved to heat the gas to high 
temperatures. In order for Ohmic dissipation to become effective, the gas should be pre-heated, which could
be achieved by shocks. Direct magnetic-field measurements in the jets of DG Tau would be important to
quantitatively assess the heating mechanism, e.g., based on radio synchrotron emission
from accelerated electrons \citep{curiel93, ray97}.

\subsection{The gas-to-dust ratio in the outer disk of DG Tau}

The hardness difference between the forward and the counter jet is easily explained by the
presence of a gas disk that absorbs the softest photons from the counter jet (see Fig.~\ref{model}). 
This observation therefore in principle offers the opportunity to measure the gas-to-dust ratio in a circumstellar 
disk (at distances of a few hundred AU) from {\it differential} absorption and extinction measurements. 
The spectral-fit results summarized in Table~\ref{jet} show a difference between the gas column densities of the two jets
of $\Delta N_{\rm H} \approx 2.7\times 10^{21}$~cm$^{-2}$, albeit with large errors. For standard assumptions 
pertaining to the interstellar medium (e.g., dust grain size distribution, atomic gas absorption with ``solar'' 
elemental composition), $N_{\rm H} \approx 2\times 10^{21}~A_{\rm V}$~[cm$^{-2}]$ \citep{vuong03}. Therefore,
$\Delta N_{\rm H}$ corresponds to a visual extinction difference of $\Delta A_{\rm V} \approx 1.4$~mag (with a 1$\sigma$ range of
$\approx$ 0.8--2.1~mag). The difference between the visual extinctions of the two jets is also not precisely known; 
\citet{lavalley97} estimate a difference of about 3~mag at a distance of 1\arcsec\ from the star, which would suggest 
nearly standard gas-to-dust ratios when combined with $\Delta N_{\rm H}$, while \citet{pyo03} propose a 
difference of as much as 14.2~mag at the same position. Better determinations of differential gas absorption 
and visual extinction are needed;  we also note that most of the jet-related X-ray counts were collected 
from larger stellar distances (2\arcsec--5\arcsec) than the distances to which the $\Delta A_{\rm V}$ values refer.

\subsection{The hard stellar source}

Because the hard
component occasionally flares, in one case being preceded by U band emission 
as in solar and stellar flares \citep{guedel07b}, it is most straightforwardly 
interpreted as coronal or ``magnetospheric''. The high temperatures are also not consistent with shock 
heating given gas flow velocities of no more than a few 100~km~s$^{-1}$ (as observed in the jets, or 
inferred from free-fall onto the stellar surface). 

The hard component of the DG Tau point source is unusually strongly absorbed, with 
$N_{\rm H} \approx 2\times 10^{22}$~cm$^{-2}$.
From the stellar extinction, $A_{\rm V} = 1.5-3$~mag (\citealt{guedel07b} and references 
therein), we expect $N_{\rm H} \approx (3-6)\times 10^{21}$~cm$^{-2}$. We therefore find from 
Table~\ref{spec}, that the dust content in the accretion streams is depleted by factors of 3--6.

The excess photoelectric absorption
requires the presence of gas that is depleted of dust in order to suppress optical extinction.
One possibility are relatively cool stellar or disk winds. However, as they expand to large stellar distances,
they would also affect the jet components and the soft ``stellar'' component unless these were placed
very far from the star. Also, it is not clear why such winds would be dust-depleted, or if they initially are, why
they would not form dust.

Dust is destroyed at several stellar radii due to stellar irradiation; for DG Tau in particular,
\citet{guedel07b} estimated the dust sublimation radius at (7--10)$R_*$, similar to the corotation radius
(see below). The region inside the corotation radius is thought to be dominated by magnetic accretion.
The excess absorption is thus most easily explained by the infalling, dust-depleted massive accretion
streams. These observations therefore provide indirect evidence for dust-depleted accretion streams that
absorb the X-ray emission from the underlying corona. 

We now show that the amount of photoelectric absorption is plausible. Estimating the gas column density
along the line of sight through the accretion streams to the stellar corona requires knowledge
of the accretion geometry, which is not available. As a conservative limit, we assume spherically
symmetric infall. The corotation radius, at which the mass begins to accelerate toward the star, can 
be derived from Kepler's law:
\begin{equation}
a_{\rm corot} = \left( {GM_*P^2\over 4\pi^2}\right)^{1/3} = 9.2\times 10^{11}~{\rm cm} = 5.6R_*
\end{equation}
where $M_* = 0.91M_{\odot}$ is the stellar mass \citep{briceno02}, $R = 2.46R_{\odot}$ is the
stellar radius \citep{guedel07a}, $P = 6.3$~d is the rotation period \citep{bouvier93},
and $G = 6.673\times 10^{-8}$~dyn~cm$^2$~g$^{-2}$ is the constant of gravitation. 
For a semi-circular funnel stream guided by magnetic fields, the highest elevation is reached 
midway between the corotation radius and the stellar surface, at $r = 3.3R_*$. There, the stream's
velocity, $v_{\rm in} = 1.3\times 10^7$~cm~s$^{-1}$ if free-fall acceleration is assumed (as an upper limit)
starting from the corotation radius, is about 40\% of the fall velocity at the stellar surface, 
and the ionization degree of the gas may still be moderate. We note that the jet-axis inclination
and by inference the most likely stellar rotation-axis inclination is $\approx$38~deg \citep{eisloeffel98}.
For an order-of-magnitude estimate, 
we assume spherical infall at this radius, over a radial distance of 1$R_*$. The mass accretion rate in this
case is, for spherical symmetry
\begin{equation}
\dot{M} \approx 4\pi r^2  n_{\rm H}m_{\rm p}v_{\rm in} \approx 8.9\times 10^7 n_{\rm H}
\end{equation}
where $n_{\rm H}$ is the hydrogen number density, and $m_{\rm p}$  is the proton mass; we assume a
mean  mass per particle of $m_{\rm p}$. The observed accretion rate is 
$\dot{M} = (10^{-7.34}-10^{-6.13})M_{\odot}$~yr$^{-1}$ \citep{white01, white04}.
We thus find
$n_{\rm H} \approx 3.3\times 10^{10}$~cm$^{-3}-5.3\times 10^{11}$~cm$^{-3}$. Integrated over one stellar radius,
the absorption column density is $N_{\rm H} \approx 5.7\times 10^{21} - 9.1\times 10^{22}$~cm$^{-2}$,
in agreement with the measured values of order $2\times 10^{22}$~cm$^{-2}$.

More confined or slower accretion streams will produce higher densities, while ionized gas will reduce
photoelectric absorption. We emphasize that the high accretion rate is crucial to obtain high
$N_{\rm H}$ in the above estimates. All TAX sources reported before \citep{guedel07b} are indeed
very strongly accreting T Tauri stars.

\subsection{The soft ``stellar'' source}

The soft emission commonly seen in X-ray spectra  
from other T Tauri stars, formed at a few MK as part of a wide distribution of plasma in the magnetic corona
(e.g., \citealt{preibisch05}), is not detected here but is absorbed, leaving only a ``hard'' coronal component
above 1.5~keV in our spectra. In contrast, we see a very strong, separate soft component that is not related to the
coronal spectrum as judged from its very different absorption and the absence of variability during
flares seen in the coronal component at higher photon energies \citep{guedel07b}.
The strong absorption of the coronal X-ray component makes an origin of the soft {\it spectral} emission from a 
location close to the stellar surface unlikely. In fact, our new observations  suggest that the gas absorption of the soft X-ray component,
$N_{\rm H} = 1.3~(0.7-2.4)\times 10^{21}$~cm$^{-2}$ (90\% error range; Table~\ref{spec}),
is {\it lower} than the absorption suggested from the visual extinction to the star, $N_{\rm H}(A_{\rm V}) 
\approx (3-6)\times 10^{21}$~cm$^{-2}$ (based on $A_{\rm V} \approx 1.5-3$~mag, see 
\citealt{white01}, \citealt{white04},  \citealt{muzerolle98}, and \citealt{hartigan95}). A likely
origin of these X-rays is the base of the forward jet. Such an origin is suggested by i) the unusually
soft emission not usually seen in T Tauri stars \citep{guedel07a}, ii) the low $N_{\rm H,1}$, and
iii) the explicit evidence for jets in the {\it Chandra} image. Further, $N_{\rm H,1}$ and $kT_1$
agree with the corresponding values of the forward jet (Table~\ref{jet}). We therefore suggest
that the jets continue to be X-ray sources to smaller stellar distances, producing an order 
of magnitude more soft X-rays within the {\it Chandra} point-spread function of 
$\approx 2\arcsec$ radius than outside this radius.

\section{Summary and conclusions}

We have unambiguously detected a bipolar X-ray jet associated with the strongly accreting
classical T Tauri star DG Tau. This is the first bipolar X-ray jet reported from a pre-main sequence
star. The jet is extended at {\it Chandra's} spatial resolution and can be followed down into
the PSF of DG Tau itself. The jets are roughly symmetric as far as length and structure
are  concerned, reaching out to about 5\arcsec\ from the star.

The X-ray emitting gas of the jet is relatively cool, $T \approx 3.4$~MK, which however still
poses a problem for our understanding of the heating mechanism. Shock velocities in this 
jet are too small to heat gas to such temperatures. Ohmic heating by magnetic-field
driven, dissipating currents may be an alternative.

We find various gas absorption columns toward the four X-ray components detected in the
DG Tau system.  The soft stellar component shows $N_{\rm H}$ smaller 
than the value derived from the stellar visual extinction assuming standard gas-to-dust 
ratios. This suggests that the soft spectral emission originates from a region ``in front'' of the star,
i.e., we identify the soft component with X-ray emission similar to that from the forward jet 
but produced too close to the star to be resolved in the {\it Chandra} images.

In contrast, the counter jet suggests
stronger absorption, which is expected because its X-rays traverse the  extended outer gas disk.
A determination of the gas-to-dust ratio is therefore in principle possible by measuring
the {\it differential} absorption and extinction of the two jets. We have succeeded in
determining the difference in $N_{\rm H}$ although not the absolute values of each
column density. The gas column of the intervening extended disk structure is relatively
small ($2.7\times 10^{21}$~cm$^{-2}$), corresponding to 1.5--2 magnitudes of visual extinction
for standard gas-to-dust ratios. An extinction difference of 3~mag has been reported in
a previous paper \citep{lavalley97}, i.e., the outer-disk composition is nearly compatible
with interstellar gas-to-dust ratios. However, more accurate determination both of the 
absorption difference and the visual extinction difference is needed.
 
Finally, the hard component in the stellar spectrum is attributed to a (flaring) corona.
The excess photoelectric absorption is ascribed to accretion gas streams that are dust depleted because
the dust destruction radius is similar to the radius of the inner disk edge, thought to be close
to the corotation radius.

Why is the detection of X-ray jets important? The combined power of the resolved jets and the
unresolved soft spectral component is of order $10^{29}$~erg~s$^{-1}$ or similar to 
the X-ray output of a relatively X-ray faint T Tauri star. This emission is distributed
above the inner accretion disk.  It is therefore an important contributor to X-ray heating
and ionization of gaseous disk surfaces \citep{glassgold04}, a role that has been studied
earlier in the case of active galactic nuclei (in the context of the ``lamppost model'', 
see, e.g., \citealt{nayakshin01}).
We speculate that protostellar jets in general develop the same kind of jet X-ray emission,
but these sources remain undetected close to the star because of strong photoelectric
absorption. In those cases, jet X-rays may act as a dispersed ionization source to affect a 
larger volume of gas than the stellar coronal source alone.

\begin{acknowledgements}
M.~A. acknowledges support from a Swiss National Science Foundation Professorship
(PP002--110504). M.~A. and S.~S. acknowledge support by NASA through CXC award SAO
GO6-7003. The CXC X-ray Observatory Center is operated by the Smithsonian Astrophysical
Observatory for and on behalf of the NASA under contract NAS8-03060.
\end{acknowledgements}

\end{document}